# Image registration of an electromagnetic tracking enabled afterloader and CT using a phantom for the quality control of implant reconstruction

Isaac Neri Gomez-Sarmiento[1,2], Daline Tho[1,2], Christopher Dürrbeck[3,4], Wim de Jager[5], Denis Laurendeau[6], Luc Beaulieu[1,2]

[1] Département de physique, de génie physique et d'optique, et Centre de recherche sur le cancer, Université Laval, Québec, Québec, Canada.

[2] Service de physique médicale et de radioprotection, Centre Intégré de Cancérologie, CHU de Québec – Université Laval et Centre de recherche du CHU de Québec, Québec, Québec, Canada.

[3] Department of Radiation Oncology, Universitätsklinikum Erlangen,Friedrich-Alexander-Universität Erlangen-Nürnberg (FAU),Erlangen,Germany.

[4] Comprehensive Cancer Center Erlangen-EMN (CCC ER-EMN), Erlangen,Germany.

[5] Elekta Brachytherapy, Veenendaal, The Netherlands.

[6] Département de génie électrique et de génie informatique, Faculté de sciences et de génie, Université Laval, Québec, Québec, Canada.

**Correspondence**

Luc Beaulieu, Service de physique médicale et de radioprotection, Centre Intégré de Cancérologie, CHU de Québec – Université Laval et Centre de recherche du CHU de Québec, Québec, Québec, Canada.

Email: luc.beaulieu@phy.ulaval.ca

## Abstract

**Background:** Electromagnetic tracking (EMT) has been researched for brachytherapy applications, showing a great potential for automating implant reconstruction, and overcoming image-based limitations such as contrast and spatial resolution. One of the challenges of this technology is that it does not intrinsically share the same reference frame as the patient's medical imaging. **Purpose:** To register the reference frame of an EMT-enabled brachytherapy afterloader with the reference frame of a CT scan image of a rigid phantom for the quality control of EMT-based implant reconstruction. **Materials/Methods:** Eleven 6F-catheters (20 cm long), one 6F round tip catheter (29.4 cm long) and a tandem and ring gynecological applicator (Elekta, CT/MR 60°, 40 mm long tandem, 30 mm diameter ring) were placed in a rigid custom-made phantom (Elekta Brachytherapy, Veenendaal, The Netherlands) to reconstruct their geometry using a five-degree of freedom EMT sensor attached to an afterloader's check cable. All EMT reconstructions were done in three different environments: disturbance free (no metal nearby), CT-on-rails brachytherapy suite and MRI brachytherapy suite. Implants were placed parallel to a magnetic field generator and were reconstructed using two different acquisition methods: step-and-record and continuous motion. In all cases, the acquisition is performed at a rate of approximately 40 Hz. A CT scan of the phantom inside a water cube was obtained. In the treatment planning system (TPS), all catheters in the CT images were manually reconstructed and the applicator reconstruction was achieved by manually placing its solid 3D model, found in the applicator library of the TPS. The Iterative Closest Point and the Coherent Point Drift algorithms were used, initialized with four known points, to register both EMT and CT scan reference frames using corresponding points from the EMT and CT based reconstructions of the phantom, following three approaches: one gynecological applicator, four interstitial catheters inside four calibration plates having and S-shaped path, and four 5 mm diameter ceramic marbles found in each of the four calibration plates. Once registered, the registration error (perpendicular distance) was computed. **Results:** The absolute median deviation from the expected value for EMT measurements in the disturbance free environment, CT-on-rails brachytherapy suite and MRI-brachytherapy suite are 0.41 mm, 0.23 mm, 0.31 mm, respectively, while for the CT scan it is 0.18 mm. These values significantly lie below the sensor's expected accuracy of 0.70 mm ($p < 0.001$), suggesting that the environment did not have a significant impact on the measurements, given that care is taken in the immediate surroundings. In all three environments, the two acquisition and three registration approaches have a mean and median registration errors that lie at or below 1 mm, which is lower than the clinical acceptable threshold of 2 mm. **Conclusions:** This work explored various EMT-image registration approaches for interstitial catheters in GYN brachytherapy. It was demonstrated that a registration based on a GYN tandem and ring applicator geometry, reconstructed with EMT and the TPS solid applicator library, allows mean registration errors within clinically acceptable accuracy, comparable to CT-based reconstruction but within a few minutes.

## 1. Introduction

In modern high dose-rate brachytherapy (HDR BT), after the implants are placed in or on the patient, 3D medical images are acquired to contour the target and organs at risk (OAR) and to localize the 3D coordinates of all channels where the source will move through; this latter step is also called implant reconstruction. This information is used for dose optimization, by finding the optimal dwell positions and dwell times for each patient-specific geometry. The knowledge of the position of the source inside the implants relative to the patient anatomy, is critical and could otherwise result in an overdosage of the OARs or underdosage to the target volume.

Electromagnetic tracking (EMT) technology has been proposed for brachytherapy applications such as the reconstruction of the 3D geometry of implants, quality assurance and error detection.[1-9] An EMT sensor placed within a low intensity, varying magnetic field produced by a field generator (FG), provides positions and orientations within the FG detection volume.[10] The idea of integrating an EMT sensor attached to an afterloader's check cable was first published by Bert et al.[11]. Kallis et al.[12] described a prototype afterloader system used to detect and quantify changes and uncertainties in the implants, stating that the disadvantage of the EMT-enabled afterloader is that all the coordinates are measured in the EMT reference frame and not in the 3D medical image reference frame. Beaulieu et al.[13] proposed an approach for US-based prostate brachytherapy where multiple sensors and a specific image-EMT calibration procedure was devised. However, to this day, no specific approach was proposed for CT scan in the context of gynecological (GYN) brachytherapy. Hence, methodologies for registration of EMT and CT scan reference frames must be found if these coordinates are to be used for treatment planning purposes along with the contouring of the target and OARs.

In this study, we are proposing a rigid phantom design that can be used to perform the registration of the catheter/applicator reconstruction made by an EMT-enabled afterloader and CT scan. The objective is to develop methodologies to bring together both coordinate systems, enabling automated channel reconstruction and potentially other tasks part of an EMT quality control (QC) program. The phantom was used to test a registration method that leverages modern treatment planning system (TPS) ability to place a pre-reconstructed applicator (applicator library model) within an acquired 3D imaging data for QC in brachytherapy.

## 2. Materials and methods

### 2.A Research afterloader unit

A research prototype of brachytherapy afterloader based on the Flexitron (Elekta Brachytherapy, Veenendaal, The Netherlands) was used in this work. It has a custom-built check cable that integrates a 5-DOF cylindrical EMT sensor (0.8 x 4.0 mm). It records position (3-DOF) and orientation (2-DOF) at a rate of approximately 40 Hz, when it is nearby a FG (NDI Aurora V3), whose detection volume is 50 x 50 x 50 cm$^3$ and effective detection volume of 30 x 30 x 30 cm$^3$.[14] More details on this prototype afterloader system can be found in Kallis et al.[12]

### 2.B Phantom

Figure 1a shows the phantom used in this study. It is composed of a 30 x 30 x 30 cm$^3$ plastic frame compatible with CT, X-ray and MR scanners, also called Pre-Treatment Brachytherapy Verification (PTBV) QC tool. Multiple plastic inserts can be used with this phantom. First, inside four template plates with a total size of 150 x 100 x 46.16 mm$^3$ (Fig. 1b), seven 20 cm long 6F interstitial catheters (lumen diameter: 1.4 mm) were placed in straight and bent paths (Fig. 1c); Second, a clamp to hold a gynecological applicator (here a tandem and ring were used - lumen dimeter: 3 and 2.5 mm, respectively), where a 29.4 cm long 6F round catheter (lumen diameter:1.4 mm) was placed in one of the applicator's holes (Fig. 1d); Finally, four calibration plates (50 x 25 x 5 mm$^3$), also called sensor verification tools, at different fixed positions (Fig. 1e). These calibration plates have an S-shaped path inside which 6F catheters can be placed and contain a 5 mm diameter ceramic marble that serve as reference points for imaging. The CAD models of the PTBV phantom and calibration plates were available for comparison purposes.

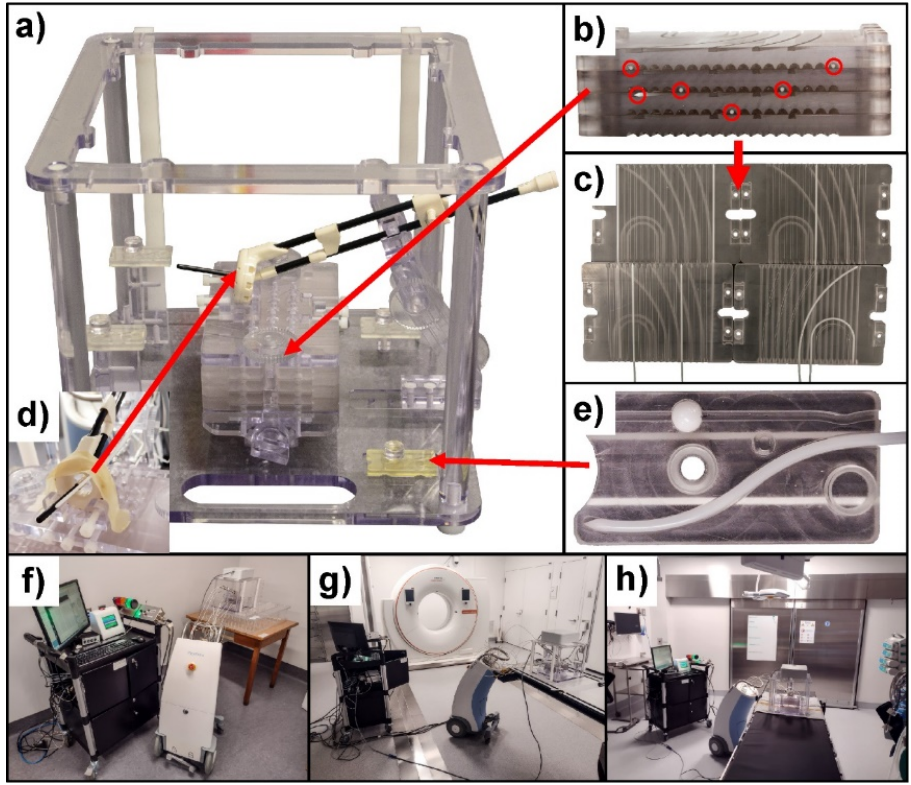

**Figure 1.** Details of the phantom and environments where EMT reconstruction were performed. a) Phantom. b) Plastic template plates. c) Seven 6F interstitial catheters placed inside the template plates. d) Rounded tip catheter inserted in one of the applicator's holes. e) Four calibration plates with 6F interstitial catheters inside and ceramic marbles (white spheres). f) Disturbance free environment (no metal nearby). g) CT-on-rails brachytherapy suite. h) MRI brachytherapy suite (outside the Faraday cage door).

## 2.C EMT and CT scan data acquisition

For the EMT measurements, the FG was placed on top of the phantom within its detection volume, such that the interstitial catheters direction is parallel to it.[14] Due to its working principle relying on electromagnetic induction, EMT is known to be susceptible to disturbances of the magnetic field.[1,2,15,16] To account for this, all EMT reconstructions were done in three different environments: a disturbance free environment (no metal nearby), in a CT-on-rails brachytherapy suite and an MRI brachytherapy suite (Fig. 1f-h). Fourteen transfer tubes were used to connect the implants to the afterloader. The implant geometry was reconstructed using two data acquisition techniques already used by Tho et al.[17]: step-and-record and continuous motion. For the step-and-record technique, the EMT sensor was set to move by 1 mm steps in curved interstitial catheters and ring and 2 mm steps for straight interstitial catheters, round catheter and tandem. In all cases, a recording (dwell) time of 1 s with approximately 40 sample per second (40 Hz) was used.

After the EMT sensor reaches the most distal dwell position of the implants, the EMT sensor starts to retract towards the afterloader, and at this point the continuous motion technique takes place, and the sensor moves backward at a constant speed of 2.5 cm/s still acquiring positional data at the same sampling rate. The speed of the sensor is not the same as the usual clinical speed of the check cable. The speed was reduced to get a higher resolution because the sampling rate cannot be increased.

For the CT scan acquisition, the phantom was placed in a plastic cube (40.5 x 32.2 x 39 cm$^3$) filled with water to improve the contrast of the implants, and was scanned using a SOMATOM Confidence (SIEMENS Healthineers, Forchheim, Germany), following a clinical protocol for GYN brachytherapy: voxel size of 0.46 x 0.46 x 2 mm, 337.5 mAs eff. and 120 kVp. An iterative metal artifact reduction (IMAR) filter was used to improve the reconstruction of the ceramic marbles.

*2.D Electromagnetic tracking data processing and CT scan data processing*

To process the raw EMT data from the log files, an in-house code was written in Python 3.8. For the stop-and-record acquisition, at each dwell position (~40 data sample acquired in 1 second), the centroid was computed using the recorded coordinates, then an interquartile range method for outlier removal was used based on the distances of the centroid and each of the sensor positions, and the centroid was computed again without outliers. In the case of the continuous motion acquisition, a moving average filter with a window size of 4 was used to smooth the point cloud, followed by a resampling to match the distance between two contiguous dwell positions as in the stop-and-record acquisition (either 1 mm or 2 mm).

The CT-based manual reconstructions of catheters, as well as the manual placement of the CAD model of the GYN applicator (TPS solid applicator library), were done using a research version of the Oncentra® Brachy TPS, version 4.6.0 (Elekta Brachytherapy Solutions, Veenendaal, The Netherlands). After all channels were manually reconstructed, dwell positions were activated according to the same sensor dwell positions used for EMT-based automatic reconstructions.

*2.E EMT to CT scan reference frame registration methodologies*

To register EMT and CT scan reference frames, there must be corresponding point clouds in both frames. The Iterative Closest Point (ICP)[18] and Coherent Point Drift (CPD)[19] algorithms were used one after the other to achieve a rigid registration, which find optimal rotation and translation parameters, while minimizing a distance criterion. The registration was initialized with four known points from EMT and CT scan frames. The metric used to evaluate the registration methodologies is the perpendicular distance, referred as the registration error. This metric has also been used by Dürrbeck et al.[20] as a quantitative measure of geometric similarity between EMT and CT based reconstructions. In this research, three registration methodologies were used based on the different point clouds obtained from the reconstructions of the phantom: 1) Applicator-based registration, 2) S-shaped plate-based registration, 3) marble-based registration.

In the applicator-based registration, the GYN applicator (tandem and ring) was fully reconstructed by the EMT sensor. Its corresponding point cloud in the CT reference frame was determined as explained in section 2D. Four-point initialization was done using first and last dwell position from ring and tandem channels, respectively, in both reference frames.

In the S-shaped plate-based registration, the EMT point clouds are from all four catheters inserted in the calibration plates. The portion of the catheters outside the calibration plates was not considered as it is not sufficiently rigid for registration purposes. Corresponding point clouds in CT scan reference frame were chosen from the manual reconstruction done in the TPS. Four-point initialization was done using the centroid of each of the four S-shaped point clouds in both reference frames.

In the marble -based registration, the position of the marble in the EMT reference frame could not be directly determined by placing an EMT sensor inside the centroid of the marble. An indirect method to obtain this position is to use the CAD model of the phantom, specifically the middle path of the S-shaped path and the centroid of the marble inside the four calibration plates. The geometry of the CAD model of the calibration plate was validated by comparing its 3D optical scanned counterpart, for each of the 4 calibration templates, using a MetraScan 210 (Creaform, Lévis, Canada) laser scanner. A rigid registration was applied to align the CAD S-shaped point cloud to its EMT counterpart, providing the centroid of the marble in the EMT reference frame. To get the centroid of the marbles in the CT scan reference frame, a least-squares sphere fitting was applied to the point cloud obtained from the edge pixel coordinates of the marbles on each CT slice. These coordinates were obtained by applying a thresholding technique followed by a contour detection algorithm[21] implemented using the OpenCV library.

## 3. Results

Figure 2a shows an example of the readings made by the EMT sensor at a given dwell position. The green dot represents the centroid of the point cloud taking into consideration all coordinates recorded in 1 second and the red dot is the centroid of the point cloud with outliers left out (black lines were drawn between the red dot and the inliers). The EMT step-and-record reconstruction of all catheters and GYN applicator, number of points per channel, as well as the expected distance between two consecutive points in a channel are described in Figure 2b. Figure 2c provides a visual comparison of the 5 parallel catheters arranged in V-shape in transversal view (see also Fig. 1b), reconstructed with EMT and its ground truth. Each catheter was reconstructed with 48 points evenly spaced by 2 mm, but to improve visualization it was downsampled. Figure 2d shows boxplots superimposed with violin plots to show the full distribution of the absolute interpoint distance differences between the ground truth reconstruction of the catheters compared to manual CT and automatic EMT-based reconstruction (28,680 possible distance differences for 240 points); for the CT-based reconstruction, the median is 0.18 mm, while the mean is 0.24±0.22 mm; for the EMT-based reconstruction in disturbance-free environment, the median is 0.41 mm, while the mean is 0.47 ±0.33 mm.; for the EMT-based reconstruction in CT-on-rails brachytherapy suite, the median difference to ground truth is 0.23 mm, while the mean is 0.29 ±0.24 mm; finally, for the EMT-based reconstruction in MRI brachytherapy suite, the median value is 0.31 mm, while the mean is 0.37 ±0.28 mm. A Wilcoxon signed rank test showed that for each distribution, these medians significantly lie below 0.70 mm ($p<0.001$), which is the EMT sensor accuracy as specified by the vendor.

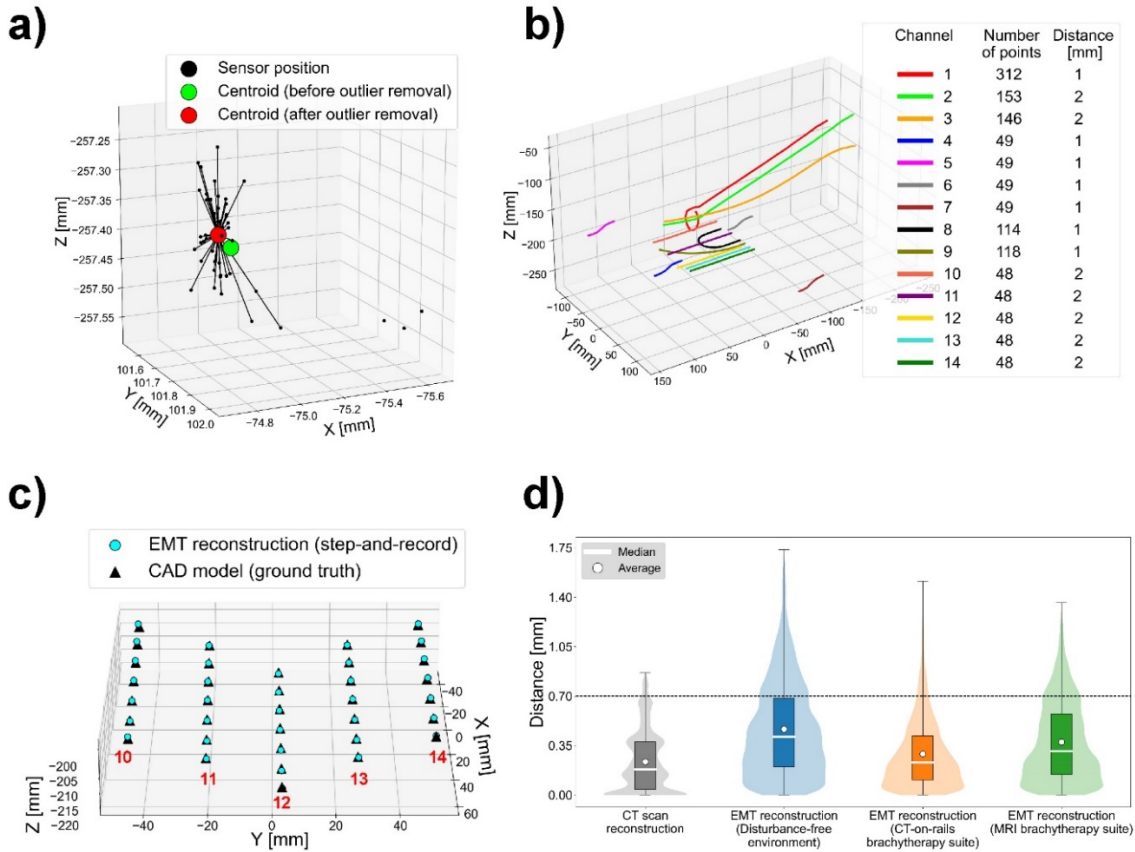

**Figure 2.** Electromagnetic tracking reconstruction of the phantom and comparison of channels 10-14 with respect to CAD model (considered as ground truth). a) EMT measurements in a dwell position of a particular channel. b) Step-and-record EMT reconstruction of all fourteen channels of the phantom; the distance column in the table is the expected distance between two consecutive points in a channel. c) 3D plot of the reconstruction of five 6F catheters arranged in V-shape inside template plates. d) Interpoint distance differences computed with all possible distances between points from CT-based or EMT-based reconstructions of channels 10-14, subtracted from the CAD model distances (used as an absolute reference). The line at 0.70 mm represents the sensor accuracy as specified by vendor.

One CT scan slice of the calibration plate is given in Figure 3a. The white dots correspond to the CT-based manual reconstruction of the S-shaped catheter, while red points belong to the edge of the marble and its centroid. Figure 3b depicts the fitted spheres to the coordinates of the marble's edge; for the inner sphere, the fit was done knowing the radius (2.50 mm), while for the outer sphere, the fit was done having the radius as fitting parameter. The mean error between the ground truth radius (2.5 mm) and the estimated radius by sphere fitting is 0.79±0.06 mm.

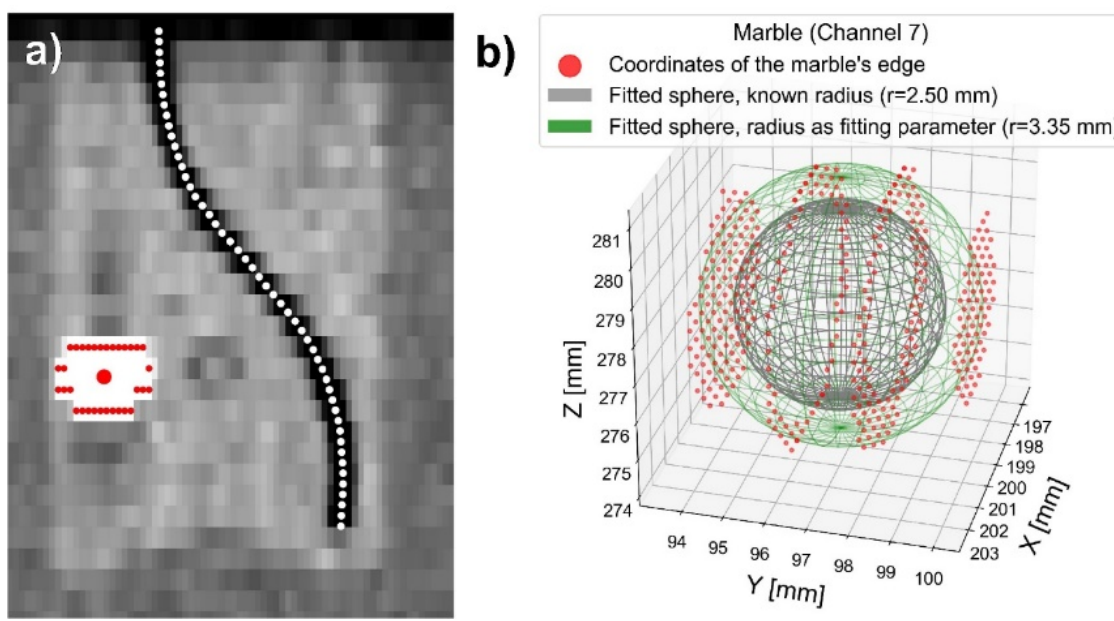

**Figure 3.** Determination of the centroid of the marbles within the calibration plate. a) CT scan of the middle slice of the S-shape template overlaying in red dots the edges of the ceramic marble and its centroid and in white dots the middle path of the S-shape cavity. b) Sphere fit of the edge pixels of the marble with known radius (inner sphere) and radius as fitting parameter (outer sphere).

Figure 4 depicts some views of the step-and-record EMT reconstructions of the phantom registered to the CT scan image reference frame. These reconstructions were performed in the MRI brachytherapy suite (outside the Faraday cage), where the phantom was placed on the operating table as shown in Figure 1h. The blue dashed line represents the EMT reconstruction registered with the applicator, while the green continuous line represents the same reconstruction but registered with the reference marbles.

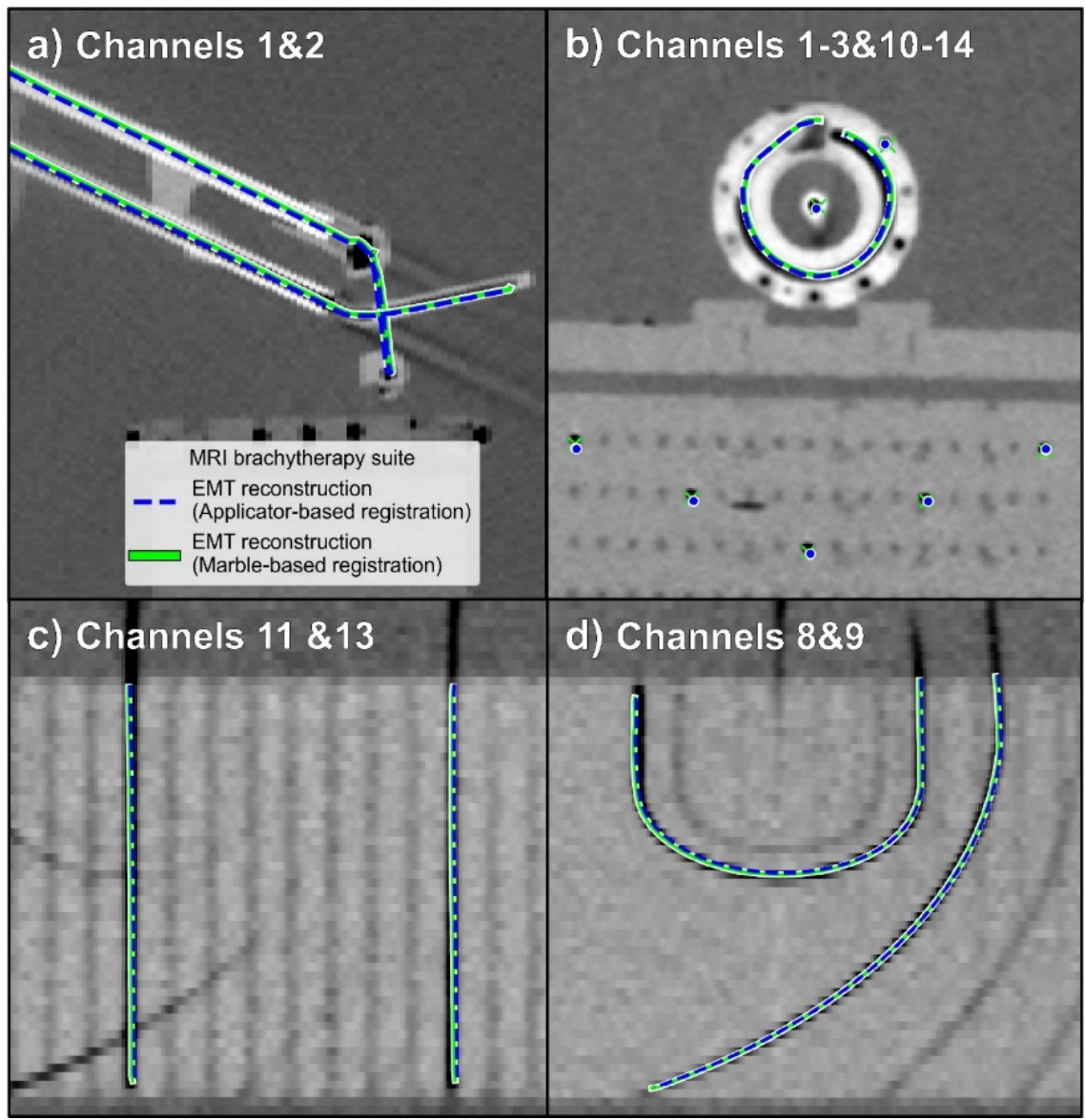

**Figure 4.** EMT reconstructed implants registered in the CT scan reference frame. The dashed line represents the applicator-based registration reconstruction while the full line corresponds to the reference marble-based registration: a) Sagittal view of the applicator. b) Axial view of the applicator, round tip catheter and 5 parallel interstitial catheters arranged in V-shape. c) Coronal view of 2 parallel catheters. d) Coronal view of 2 bent catheters.

EMT to CT scan registration error for individual channels in the three different environments, for step-and-record EMT reconstructions are presented in the form of boxplots in Figure 5. Applicator-based and marble-based registration methodologies were used for Figure 5a-b, respectively. It can be noticed that in both figures, the median and mean registration error, as well as the third quartile, of all individual channels are below 2 mm. The largest differences being for channels that are the farthest from the FG (i.e. channel 7).

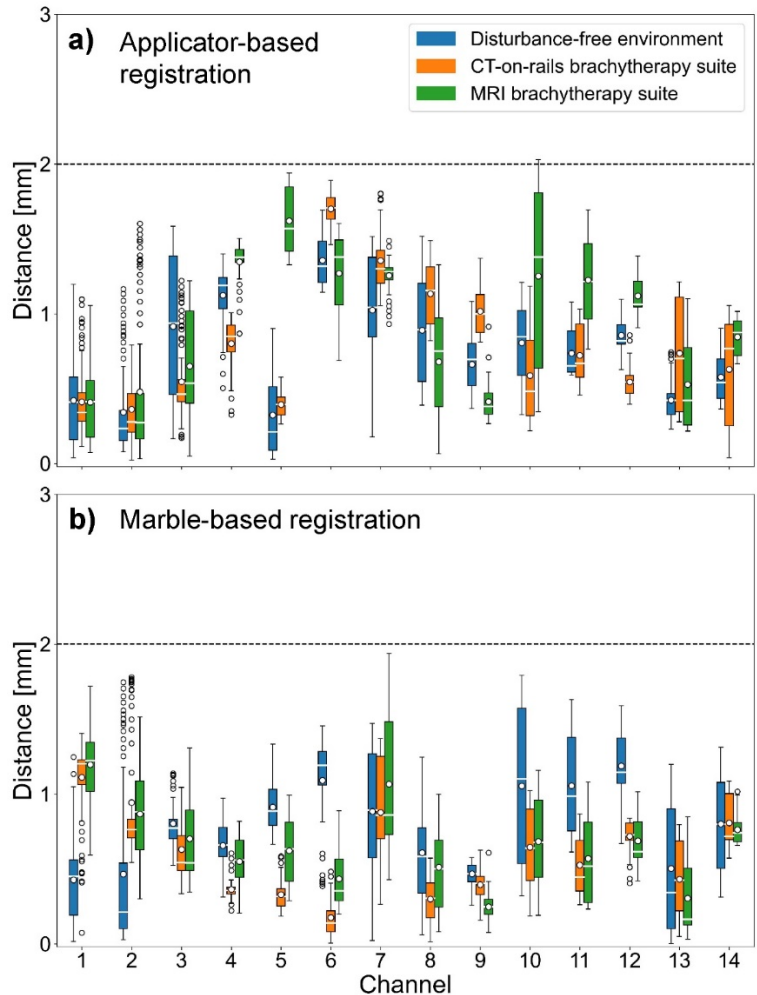

**Figure 5.** EMT to CT scan registration errors for individual channels for step-and-record acquisition. The horizontal line at 2 mm is the clinical acceptable threshold. The white line and dot on the boxplot represent the median and mean, respectively. a) Applicator-based registration. b) Marble-based registration.

Finally, the overall registration error for each registration methodology, acquisition method and environment are presented in Figure 6. The maximum values are smaller for the step-and-record acquisition (Fig. 6a) compared to continuous motion acquisition (Fig. 6b). Nonetheless, the mean and median are all at or below 1 mm, while the third quartile values are well below 2 mm.

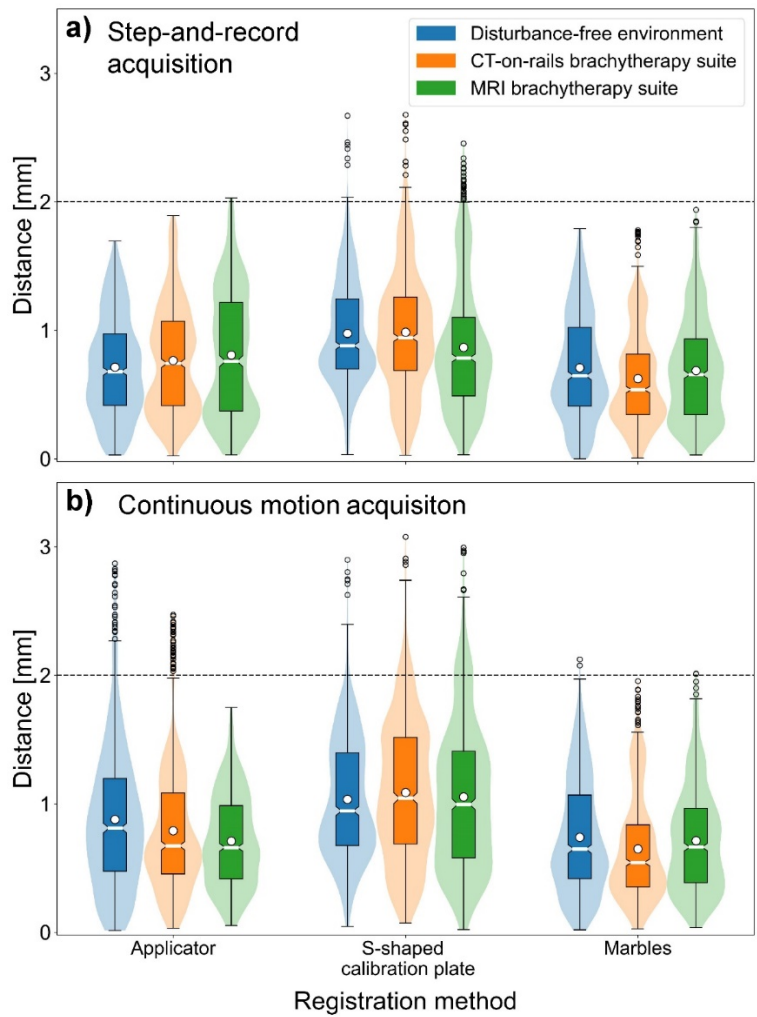

**Figure 6.** EMT to CT scan overall implant registration errors for the three studied registration methods: applicator-based, s-shaped calibration plate and marble-based, each in the three environments explored. The horizontal lines at 2 mm represent the clinical acceptable threshold. The white line and dot on the boxplot represent the median and mean, respectively. a) Step-and-record acquisition. b) Continuous motion acquisition.

## 4. Discussion

In this study it was possible to register the reference frame of an EMT-enabled afterloader into a CT scan reference frame. The afterloader has an EMT sensor attached to its check cable, which allowed to reconstruct the 3D geometry of catheters and tandem and ring GYN applicator inside a rigid phantom. The registration was achieved by using common points in both reference frames, such as the reconstruction point clouds from the applicator, catheters and the centroid of marbles. In the case of continuous motion acquisition, a speed of 2.5 cm/s was chosen as it is accurate enough to detect catheter shifts of 1 mm, based on the prior study of Tho et al. for the same system.[17] A higher check cable speed would result in more outliers, due to a slight asynchronism of the timestamps of the EMT system compared to the afterloader device, according to Dürrbeck et al.[22]

The absolute interpoint distance differences of the 5 parallel catheters arranged in V-shape, is a measurement of the deviation of all possible distances between points in EMT or CT reconstruction point clouds from the CAD point cloud (considered as ground truth). This metric obtained in all 3 environments result in a mean, median and third quartile values all below 0.70 mm. These values are within the sensor accuracy [7,20], as shown by the Wilcoxon signed rank test, supporting the evidence that the environment did not have a significant effect on the measurements, provided that care is taken concerning the immediate environment close to the field generator.[1,2,15,16] It's worth mentioning that in the case of the EMT measurements done on the MRI brachytherapy suite, the phantom and FG were placed on the intervention table (see Fig. 1h), where the patient would lie down. Measurements done on a gurney closer to the MRI safety door showed visual distortions specially in the farthest channels from the FG (data not shown).

One of the reasons the extracted radius of the marbles in the CT scan does not exactly match the ground truth can be explained due to the limited resolution of the CT scan (0.46x0.46x2 mm – visually the marbles do not look like a sphere on the CT image). Qualitatively, the EMT reconstruction overlaps well with the CT scan image in all views, noticing a slight offset from the center of the implants (Fig. 4b), which can be in part due to radial fluctuations as the sensor diameter (0.8 mm) is smaller than the lumen diameter of a 6F catheter (1.4 mm), tandem (2.5 mm) and ring (3 mm), with deviations from the center of the path that could reach up to ±0.3 mm, ±0.85 mm and ±1.1 mm, respectively. In the clinic, the center of the catheters is reconstructed, even if the source will not follow exactly that path, so the EMT sensor path could be another way to approximate the actual source path, but further validation would be needed.

When comparing registration error for individual channels, it can be noticed that the registration method influences their distribution. For example, when doing registration using only the applicator (channel 1 and 2), those two channels have lower errors compared to a marble-based registration. If using the marble-based registration method, which makes use of channels 4-7, these four channels have lower errors compared to an applicator-based registration. There are multiple explanations for these differences. First, because of the phantom design, the clamp that holds the applicator allows for more degrees of freedom, but care must be taken that the transfer tube does not induce motion of the applicator (the clamping is not perfectly rigid). Conversely, the calibration plates fit perfectly within the phantom geometry. Second, the applicator has an internal lumen diameter greater than a 6F catheter, such that the check cable traveling in it has more freedom for radial motions, thus increasing the positional uncertainties. Third, the positional errors within the FG detection volume are not homogenous as demonstrated by Boutaleb et al.,[14] increasing when the sensor is farther from the FG, thus the overall accuracy of the registration depends on the spatial distribution of the reconstruction point clouds over the FG detection volume. Finally, in applicator-based registration, there is a user-dependent variability related to the manual placement of the applicator library model in the CT scan image set, whereas in marble-based registration the marble centroids in CT scan are determined automatically using image processing operations. The interobserver variability of tandem and ring manual reconstruction, which takes between 3-5 min to reconstruct manually, has already been studied by Hrinivich at al.[23], where they found a mean standard deviation interobserver variability of $0.83 \pm 0.31$ mm and $0.78 \pm 0.29$ mm for the ring and tandem, respectively.

When comparing overall registration errors for all registration methods, EMT acquisition methods and environments, the summary statistics are well within the clinical acceptable threshold (<2 mm) for MR-based applicator reconstruction for GYN brachytherapy according to TG303[24], resulting in average variations for dosimetric indices and DVHs between 1-2%.

More importantly, using the reference templates attached to this phantom, we were able to evaluate that the applicator-based registration method, which does not need any reference marbles or other external references, does enable a reproductible and accurate (<2 mm) registration of interstitial catheters. Using either the step-and-record or continuous motion approach, the accuracy is on the same order as the CT scan voxel size.

Using the step-and-record and continuous motion acquisition, the overall reconstruction time for all 14 channels presented in this study takes about 30 min and 6 min, respectively. This is for a sensor dwell time of 1s, dwell positions of 1 and 2 mm, forward and retraction speed of 50 cm/s and 2.5 cm/s, respectively. In our clinical practice, the estimated reconstruction time in the most difficult clinical cases usually take less than 10 min (about 13 catheters plus applicator placement). However, it is worth noticing that in the clinic, whole catheters paths are not reconstructed, contrary to what was done in this work. Furthermore, the acquisition time reported here would also have decreased if using a larger sensor dwell position spacing, like the clinical studies for the breast performed by Dürrbeck et al.[20], which used sensor dwell positions of 10 mm. In that study between 11-27 catheters used in interstitial breast brachytherapy, were reconstructed by the EMT-enabled afterloader in 15 min or less.

Finally, since the ring and tandem applicator is also visible in MRI, the applicator-based registration approach proposed here should work for an MR-only GYN brachytherapy planning. In that procedure, interstitial catheters are generally difficult to visualize, thus the proposed registration approach could eliminate the need of a CT-scan for the sole purpose of the implant reconstruction. Other advantages of an EMT-MR GYN brachytherapy workflow would be that it could overcome image-based limitations such as contrast and spatial resolution. Also, when compared to deep learning aproaches[25], it does not require a retraining if changing the MRI scanner or imaging settings. Some future work includes validating an EMT-MR workflow with the phantom, simulate patient motion and automating the placement of the applicator.

## 5. Conclusion

The main objective of this study was to develop and compare methodologies to register the reference frames of a CT scan and an EMT-enabled afterloader for GYN brachytherapy. This was used to reconstruct the 3D geometry of implants inside a rigid phantom using an EMT sensor attached to its check cable. Our results show that it is possible to achieve submillimeter mean and median registration errors, well within the clinical acceptable accuracy (<2 mm) in the clinical environments where it was tested, such as CT-on-rails and MRI brachytherapy suite. The registration methodology using the EMT reconstruction of the applicator and its solid model, from the TPS applicator library, manually placed on the CT scan, should be sufficiently accurate for clinical registration of EMT reconstructed catheters to the image reference frame during GYN brachytherapy procedures.

**Acknowledgments**

This work was supported in part by the National Sciences and Engineering Research Council of Canada (NSERC) via the NSERC Alliance Grant (ALLRP 557112-20), and by an NSERC Discovery Grants (RGPIN-2019-05038). We thank Denis Ouellet, Cédric Bélanger, Janelle Morrier, Éric Poulin, Jonathan Boivin and Sylviane Aubin for their assistance and advice.

**Conflicts of interest**

This work was partially funded by a research grant from Elekta through the NSERC-Alliance grant program.